\RequirePackage{snapshot}
\documentclass[aps,prb,twocolumn,amsfonts,showpacs,longbibliography,superscriptaddress]{revtex4-2}
\usepackage{verbatim,epsfig,amsmath,amssymb,bm,epsf,graphicx,psfrag,bbold,amsthm,amsfonts}
\usepackage[toc,page,titletoc]{appendix}
\usepackage[bottom]{footmisc}
\usepackage{hyperref}
\hypersetup{
    colorlinks=true,
    linkcolor=blue,
    citecolor=blue,
    filecolor=black,
    urlcolor=blue,
}

\usepackage[capitalize]{cleveref} 
\usepackage{enumitem}
\usepackage{soul}
\usepackage{grffile}
\usepackage{framed}
\usepackage{mathrsfs}
\usepackage{esint}
\setlist{nosep}
\usepackage{placeins}
\usepackage[all]{xy}
\usepackage{color}
\usepackage[utf8]{inputenc}
\usepackage{float}
\usepackage{natbib}
\usepackage{tikz-cd}
\usepackage{verbatim}
\usepackage{leftidx}
\usepackage[normalem]{ulem}

\usepackage{notes2bib}
\usepackage{booktabs}

\def\nd{^{\vphantom{\dagger}}}

\def\frac#1#2{{\textstyle{#1 \over #2}}}

\def\Bx{{\boldsymbol x}}
\def\Bp{{\boldsymbol p}}
\def\Bone{{\boldsymbol 1}}
\def\CQ{{\mathcal Q}}
\def\CP{{\mathcal P}}

\usetikzlibrary{arrows,decorations.pathreplacing}
\usetikzlibrary{decorations.pathmorphing}

\tikzset{snake it/.style={decorate, decoration=snake}}\usepackage[T1]{fontenc}

\usetikzlibrary{decorations.markings}

\begin{document}
	
\title{Phase space fractons}

\author{Ylias Sadki}
 	\email{ylias.sadki@physics.ox.ac.uk}
	\affiliation{Rudolf Peierls Centre for Theoretical Physics, University of Oxford, Oxford OX1 3PU, United Kingdom}
\author{Abhishodh Prakash}
	\email{abhishodhprakash@hri.res.in}
   \altaffiliation{(he/him/his)} 
    	\affiliation{Rudolf Peierls Centre for Theoretical Physics, University of Oxford, Oxford OX1 3PU, United Kingdom}
        \affiliation{Harish-Chandra Research Institute,  Prayagraj (Allahabad) 211019, India}
\author{S. L. Sondhi}
	\email{shivaji.sondhi@physics.ox.ac.uk}
	\affiliation{Rudolf Peierls Centre for Theoretical Physics, University of Oxford, Oxford OX1 3PU, United Kingdom}	
\author{Daniel P. Arovas}
	\email{arovas@physics.ucsd.edu}
	\affiliation{Department of Physics, University of California at San Diego, La Jolla, California 92093, USA}	
	
\begin{abstract}
Perhaps the simplest approach to constructing models with sub-dimensional particles or fractons is to require the conservation of dipole or higher multipole moments. We generalize this approach to allow for moments in phase space and classify all possible classical fracton models with phase-space multipole conservation laws.  
We focus on a new self-dual model that conserves both dipole and quadrupole moments in position and momentum; we analyze its dynamics and find quasi-periodic orbits in phase space that evade ergodic exploration of the full phase space.
\end{abstract}
	
\maketitle
 
\section{Introduction} 

In this paper, we present a generalization of multipole conserving classical fractons previously studied in Refs. \cite{AP_NRFractonsPhysRevB.109.054313,AP_manyfractons_PhysRevB.110.024305,babbar2025classical}.
``Fractons phases'' have been studied in recent years as phases of matter in many-body quantum systems \cite{GromovRadzihovsky2022fractonReview,PretkoChenYou_2020fracton,NandkishoreHermeleFractonsannurev-conmatphys-031218-013604} whose quasi-particle excitations, ``fractons'', have mobility restricted to a manifold of lower dimension than ambient space.
The simplest examples are systems that conserve dipole and higher multipole moments, which restrict mobility through conservation laws that couple space and charge. 
In a previous paper \cite{AP_NRFractonsPhysRevB.109.054313}, we developed classical \emph{Machian fractons}, which conserve dipole moments in position and momentum (the momentum dipole would conventionally be called the total momenta), leading to ergodicity breaking via clustering in position space. 
In Ref. \cite{AP_manyfractons_PhysRevB.110.024305}, the conservation of position-space dipole moments was generalized to multipole moments which too exhibited non-ergodic dynamics qualitatively similar to dipole-conserving systems. 

In this work, we generalize these systems to conserve arbitrary multipole moments in phase space, i.e.\ both positions \emph{and} momenta. 
We first consider the general algebra of phase-space multipoles, which leads to a classification of models. 
This allows us to eliminate a large class whose algebra is unbounded and can only result in trivial dynamics.
Next, we present a general construction of Hamiltonians conserving a number of desired multipoles.
Finally, we study the dynamics of a novel self-dual model, conserving dipoles and quadrupoles in both position and momentum. 
In contrast to the models in \cite{AP_NRFractonsPhysRevB.109.054313,AP_manyfractons_PhysRevB.110.024305,babbar2025classical}, the dynamics has quasi-periodic orbits in phase space and evades ergodicity, without any clustering.

\section{Model and symmetries}
\subsection{Multipole conserving fractons}
Let us begin with a review of classical non-relativistic fractons introduced in Refs. \cite{AP_NRFractonsPhysRevB.109.054313,AP_manyfractons_PhysRevB.110.024305}. We are concerned with non-relativistic classical systems made up of $N$ point particles living in $d$ spatial dimensions subject to the conservation of spatial multipole moments 
\begin{equation}
\CQ_m^\mu \equiv \sum_a (x_a^\mu)^m\quad. \label{eq:Ql}
\end{equation}
We work within the Hamiltonian framework where the state of the system is described by $Nd$ position $x^\mu_a$ and momentum $p^\nu_a$ coordinates. The symmetry transformations  generated by the conserved quantities in \cref{eq:Ql} are shifts of momentum coordinates by position coordinate polynomials of degree $m-1$
\begin{equation}
p^\mu_a \mapsto p^\mu_a + \sum_{I = 0}^{m - 1} \beta\nd_I\, (x^\mu_a)^I\quad. \label{eq:Ql symmetry}
\end{equation}    
Analogously to $\CQ^\mu_m$, we now consider momentum multipoles:
\begin{equation}
\CP^\nu_n \equiv \sum_a (p^\nu_a)^n\quad, \label{eq:Pi_n}
\end{equation}
which generate the symmetry transformations 
\begin{equation}
x^\nu_a \mapsto x^\nu_a + \sum_{J=0}^{n-1} \alpha\nd_J\, (p^\nu_a)^J\quad. \label{eq:Pm symmetry}
\end{equation} 
Notably, spatial translation invariance enforces conservation of total momentum $\CP_1^\mu$.
We will restrict ourselves to $d=1$ and suppress the $\mu$ index for the remainder of this paper. 
All superscripts will henceforth refer to powers. 
For higher dimensions, it is difficult to construct a rotationally invariant Hamiltonian that conserves moments $\CQ_m^\mu$ for $m>2$.
In \cite{AP_NRFractonsPhysRevB.109.054313,AP_manyfractons_PhysRevB.110.024305,babbar2025classical}, we primarily considered Hamiltonians of the form
\begin{equation}
H = \frac{1}{2} \sum\limits_{a<b} (p_a-p_b)^2 K(x_a - x_b)\quad,
\end{equation}
which conserves $H$, $\CQ_1$ and $\CP_1$.
In this work, we consider more general Hamiltonians, which conserve an arbitrary combination of multipoles in position or momentum space, defined in \cref{eq:Ql,eq:Pi_n}.

\subsection{Symmetry algebra of phase-space multipoles}

We begin by reviewing the classical Multipole algebra followed by the symmetry generators $\CP_1$ and $\CQ_m$ under Poisson brackets~\cite{AP_manyfractons_PhysRevB.110.024305}
\begin{equation}
\big\{\CQ_m\,,\,\CQ_{m'} \big\}  = 0 \quad,\quad
\big\{\CQ_m\,,\,\CP_1\big\}  = m\, \CQ_{m-1}\quad.
\label{eq:ClassicalMultipoleAlgebra}
\end{equation}

As explained in \cite{AP_manyfractons_PhysRevB.110.024305}, from \cref{eq:ClassicalMultipoleAlgebra}, the conservation of $\CQ_m$ and $\CP_1$ (spatial translation invariance) will necessarily result in the conservation of $\{\CQ_1,\ldots,\CQ_{m -1}\}$.
An analogous form of the algebra will further constrain possible fracton systems which also conserve $\{\CP_m\}$. 
To study the generalized algebra of symmetries, it is useful to define the following functions.
\begin{equation}
\Pi_{m,n} \equiv \sum_a (x_a)^m \, (p_a)^n\quad.
\end{equation}
This contains the generators defined in \cref{eq:Ql,eq:Pi_n},
\begin{equation}
\CQ_m = \Pi_{m,0}\quad,\quad\CP_n = \Pi_{0,n}\quad.
\end{equation}
It can be verified that $\Pi_{m,n}$ satisfy the following algebra
\begin{equation}
\big\{ \Pi_{m,n}\,,\, \Pi_{m' n'} \big\} = (mn' - 
m' n)\, \Pi_{m+m'-1,n+n'-1}\quad. \label{eq:general algebra}
\end{equation}
We label the symmetry algebras by $\textbf{m}$, the largest position multipole, and $\textbf{n}$, the largest momentum multipole.
Although there are an infinite number of symmetry algebras, many of them lead to trivial dynamics. 
For instance, if we consider a system with $\CQ_2 = \Pi_{2,0}$ and $\CP_3 = \Pi_{0,3}$ conserved (i.e.\ $\textbf{m} \geq 2$, 
$\textbf{n} \geq 3$), the algebra in \cref{eq:general algebra} is \emph{unbounded} and leads to the conservation of $\Pi_{k,0}$ and 
$\Pi_{0,k}$ for all integers $k > 0$. 
The dynamics is then trivially frozen, with all $x_a = \mathrm{const}$ and $p_a = \mathrm{const}$.
The conjugate case $\textbf{m} \geq 3$, $\textbf{n} \geq 2$ similarly leads to unbounded algebra and frozen dynamics. 
In the remaining cases summarized below in \cref{tab:example}, the algebra of generators is closed:

\begin{table}[h!]
\centering
\begin{tabular}{l@{\hspace{5mm}}l@{\hspace{5mm}}l}
\toprule
\textbf{m} & \textbf{n} & {Conserved quantities} \\ 
\midrule
0 & $\geq 1$ & A combination of $\CP_1$ to $\CP_\textbf{n}$ \\ 
$\geq 1$ & 0 & A combination of $\CQ_1$ to $\CQ_\textbf{m}$ \\ 
$\geq 1$ & 1 & $\CP_1$, $\CQ_1$ to $\CQ_\textbf{m}$ \\ 
1 & $\geq 1$ & $\CQ_1$, $\CP_1$ to $\CP_\textbf{n}$ \\ 
2 & 2 & $\CQ_1$, $\CQ_2$, $\CP_1$, $\CP_2$, $\Pi_{1,1}$ \\ 
\bottomrule
\end{tabular}
\caption{Closed symmetry algebras resulting in non-trivial dynamics.}
\label{tab:example}
\end{table}

From this completely general construction, we see the space of possible phase-space multipole conserving systems with non-trivial dynamics is greatly reduced. 
We summarize them below.

\smallskip

\begin{enumerate}
\itemsep0.5em
\item $\textbf{m} = 0, \textbf{n} \geq 1$: $\textbf{m}=0, \textbf{n}=1$ is a general translation-invariant system, such as the familiar Newtonian dynamics. More generally, we may consider models that only conserve specific generalized momentum moments $\CP_n$, but no moments in position. See \cref{appendix: l0 m geq 1}. 

\item $\textbf{m} \geq 1, \textbf{n} = 0$: If $\CP_1$ is \emph{not} conserved, i.e. the system is not translationally invariant, then conserving $\CQ_n$ does not automatically conserve $\CQ_1$ to $\CQ_{n-1}$. For example, we could consider a model that conserves $\CQ_1$ and $\CQ_3$ but not $\CQ_2$. 
    

\item $\textbf{m} \geq 1, \textbf{n} = 1$: These are generalized multipole conserving systems, studied in \cite{AP_manyfractons_PhysRevB.110.024305}. The Hamiltonians can be constructed to be explicitly local, and exhibit ergodicity breaking wherein particles cluster in position space and spontaneously break translation invariance. Their motion is unbounded in momentum space.

\item $\textbf{m} = 1, \textbf{n} \geq 1$: The dual cases of \cite{AP_manyfractons_PhysRevB.110.024305}, when switching momenta and positions. 
These Hamiltonians are not precisely dual, however, as we still require locality in \emph{positions} (see \cref{appendix:non local momenta hamiltonians}).

\item $\textbf{m} = 2, \textbf{n} = 2$: A new, `self-dual' case exhibiting bounded motion. We will argue that ergodicity is broken even here as the system does not explore the full phase space available to it. An edge case that does \emph{not} have $\CP_1$ (i.e.\ no translational invariance) or dipole $\CQ_1$ is also possible, conserving only $\CQ_2$ and $\CP_2$.

\end{enumerate}

\subsection{Hamiltonian constructions}

We now generalize a procedure from \cite{AP_manyfractons_PhysRevB.110.024305} to construct Hamiltonians conserving general multipole moments. 
To simplify our analysis, we will assume conservation of energy so the Hamiltonian $H$ is time-independent and is itself a constant of motion. Furthermore, we will impose translation invariance i.e. conservation of $\CP_1$.
Consider the following determinant $R$, which induces symmetric interactions between $k+1$ identical particles:
\begin{equation}
R(\{x\nd_\Gamma , p\nd_\Gamma \}) = \det \begin{pmatrix}
1 & 1 & \cdots & 1\\
p\nd_1 &  p\nd_2 & \cdots & p\nd_{k + 1}\\
x\nd_1 &  x\nd_2 & \cdots & x\nd_{k + 1} \\
x_1^2 &  x_2^2 & \cdots & x_{k + 1}^2 \\
& & \ddots & \\
x_1^{k-1} &  x_2^{k-1} & \cdots & x_{k + 1}^{k-1} 
\end{pmatrix} \quad. \label{eq:R_multipole}
\end{equation}

From the properties of the determinant, $R$ is invariant under translations \cref{eq:Ql symmetry,eq:Pm symmetry}, hence conserving $\CP_1$ and $\CQ_1$ to $\CQ_k$.

In Refs. \cite{AP_NRFractonsPhysRevB.109.054313,AP_manyfractons_PhysRevB.110.024305}, the following form of Hamiltonian was considered
\begin{equation}
H = \sum_{\Gamma_{k + 1}} R^2\big(\{\Bx\nd_\Gamma , \Bp\nd_\Gamma\}\big) \prod_{a<b} K(x_a - x_b)\quad,
\label{eq:old_Ham}
\end{equation}
where $K(\Delta x)$ is a non-negative function which decays to enforce locality, and where $\Gamma_{k+1}$ denotes $(k+1)$-tuples of particles with phase space coordinates 
\begin{equation}
\Gamma\nd_{k+1} \equiv \{x\nd_{a_1},\ldots,x\nd_{a_{k+1}},p\nd_{a_1},\ldots,p\nd_{a_{k+1}} \}\quad, 
\end{equation} 
This model exhibits ergodicity breaking, wherein particles cluster in position space and spontaneously break translation invariance.

Now consider a Hamiltonian involving $(k + 1)$-body interactions as follows:
\begin{equation}
H = \sum_{\Gamma_{k+1}}  f\big(R(\{x\nd_\Gamma , p\nd_\Gamma\})\big)\quad, \label{eq:general H}
\end{equation}
where $f(\Bx,\Bp)$ is a function that is bounded from below and decays $f \rightarrow 0$ as $R \rightarrow \infty$, to ensure a sense of locality in the Hamiltonian, i.e.\ interaction terms vanish $f\rightarrow 0$ when any two particles are sufficiently far separated in position space $|x_a - x_b| \rightarrow \infty$. 
Importantly, the models we construct here will \emph{not} be strictly local, as we shall show for $\textbf{m} = \textbf{n} = 2$. 
This non-locality leads to different dynamics than Machian fractons \cite{AP_manyfractons_PhysRevB.110.024305}, which cluster in position space.
Here we will see a different form of evasion of ergodicity, instead in phase space.

Using \cref{eq:general H}, we may define Hamiltonians that conserve $\textbf{m} \geq 1, \textbf{n} = 1$. 
Exchanging positions and momenta produces models with $\textbf{m} = 1, \textbf{n} \geq 1$.
Variations including other momenta terms are discussed in \cref{appendix:non local momenta hamiltonians,appendix: l0 m geq 1}.

\section{$\textbf{m}=\textbf{n} =2$ dynamics}
We focus now on the self-dual case $\textbf{m} = \textbf{n} = 2$, which illustrates the general features of models in \cref{eq:general H}.
Consider
\begin{equation}
R(\{\Bx\nd_\Gamma , \Bp\nd_\Gamma \}) = \det \begin{pmatrix}
1 & 1 &  1\\
p_1 &  p_2 & p_3\\
x_1 &  x_2 &  x_3 \\
\end{pmatrix}\quad. 
\label{eq:R_multipole_l2_m2}
\end{equation}
The Hamiltonian \cref{eq:general H} then conserves $\CQ_1, \CQ_2, \CP_1, \CP_2$ and $\Pi_{1,1}$.

\begin{figure*}[!ht]
    \centering
    \includegraphics[width=\linewidth]{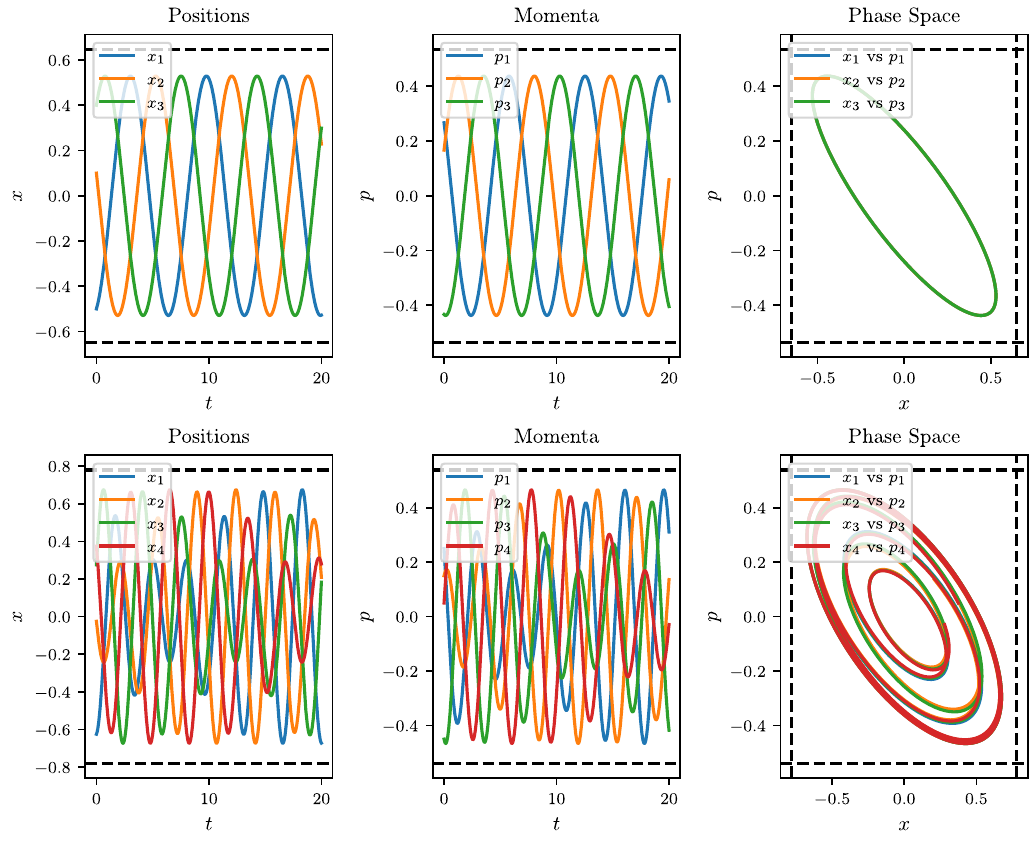}
\caption{\emph{Top}: A sample trajectory with $\textbf{m}=\textbf{n}=2$ for $N=3$ particles. \emph{Bottom}: A sample trajectory with $N=4$ particles. For both plots, dashed lines show the strict bounds enforced by $\CQ_2 = \textrm{const}$ and $\CP_2 = \textrm{const}$. These simulations use $f(R) = 1/(R^2 + 1)$.}
    \label{fig:Self_dual}
\end{figure*}

Trajectories for three and four particles are shown in \cref{fig:Self_dual}. 
The $N=3$ case is simple to solve: $H = f(R)$ is the conserved energy. Hence any function of $R$ is a constant of motion. The equations of motion for $x_i$ are
\begin{equation}
    {dx\nd_1\over dt} = f'(R) (x_2 - x_3)\quad,
\end{equation}
and cyclic permutations.
These integrate to:
\begin{equation}
    x_j = A \cos\Big( \sqrt{3}\, f'(R)\, t + \phi + \frac{2\pi}{3}j \Big)\quad,
\end{equation}
where $A$ and $\phi$ are constants fixed by the initial conditions $x_j(t = 0)$.
There are only two constants as we assume $\sum_j x_j = 0$, without loss of generality.
An analogous result follows for the momenta $p_j$. 
The motion, shown in \cref{fig:Self_dual}, involves periodic motion of all coordinates $x_i$ and $p_i$.
In particular, the trajectories of all three particles $x_i$ vs $p_i$ lie along the same ellipse, which is a bizarre feature of the highly regular dynamics.

For arbitrary number of particles $N$, the general features are similar to the $N=3$ case.
Positions and momenta of each particle oscillate close to bounds set by $\CQ_2 = \textrm{const}$ and $\CP_2 = \textrm{const}$.
In phase space \footnote{Technically we plot $x_i$ against $p_i$ for each particle on the same plot, which is not the true $2N$-dimensional phase space.}, trajectories tend to oscillate around an approximate ellipse.
Crucially in this system, the $\CQ_2$ and $\CP_2$ set strict bounds on individual $x_i$ and $p_i$.
$\CQ_2 = \sum_i x_i^2$ is conserved, hence $x_i \leq \sqrt{\CQ_2}$, and similarly $p_i \leq \sqrt{\CP_2}$ always.
The particles' trajectories tend to almost saturate these bounds. 
Importantly, these bounds are set by the initial conditions \emph{only}, and not by any length scales in the Hamiltonian.

For $N=3$, geometrical intuition provides insight into the dynamics.
$R$ in \cref{eq:R_multipole_l2_m2} is given by $R=\Bone\cdot\Bx\times\Bp$, where $\Bone = (1, 1, 1)$,
and is the sum of the components of the vector cross product $\Bx\times\Bp$, where
\begin{equation}
\Bx= \begin{pmatrix} x_1 \\ x_2 \\  x_3 \end{pmatrix} \quad,\quad
\Bp = \begin{pmatrix} p_1 \\ p_2 \\ p_3 \end{pmatrix}\quad. 
\end{equation}
Thus $R$ is simply twice the area of a triangle with coordinates $\big\{(x_1, p_1),\, (x_2, p_2),\, (x_3, p_3)\big\}$.
The geometric meaning is now clear: $R$ essentially measures how close in phase space the three particles are.
For the interaction to be local, its strength must decrease with $R$: this is why $f(R)$ must be a decreasing function.
However, if we consider surfaces of constant $R$, which is exactly constant for $N=3$ particles, one can reason that the particles may still interact with each other when $|x_1 - x_2| \rightarrow \infty$, if the momentum suitably decreases --- evidently this is at odds with the requirement of locality in \emph{position space}.
Ultimately, the constraint that prevents particles from influencing each other at arbitrary distances is the conservation law $\CQ_2 = \textrm{const}$; this sets a maximum on individual positions $x_i$.

Interestingly, trajectories in phase space for $N=3$ lie along an ellipse, despite six phase space coordinates and six conservation laws. 
Through geometric reasoning we can eliminate one of the constraints.
We begin by reformulating the conservation laws:
\begin{equation}
\begin{split}
\CQ_1 &= \Bone\cdot\Bx \qquad,\qquad 
\CP_1 = \Bone\cdot\Bp \\
\CQ_2&=  \Bx\cdot\Bx \qquad,\qquad 
\CP_2 =  \Bp\cdot\Bp \\
\Pi_{1,1} &= \Bx\cdot\Bp \qquad,\qquad
H = f(R)\quad,
\end{split}
\end{equation}
with $R=\Bone\cdot\Bx\times\Bp$.
Interpreting the first five conservation laws, the magnitudes and relative angle of $\Bx$ and $\Bp$ are fixed. 
Further, both $\Bx$ and $\Bp$ have fixed angles with the special direction $\Bone$.
The conservation of $H$ (and hence $R$) provides no additional constraints: the magnitudes and relative direction are already constants.
Hence the possible motion has one parameter, and involves the rotation of $\Bx$ and $\Bp$ in the plane perpendicular to the $\Bone$ vector: essentially, an ellipse in $x$-$p$ space.

\begin{figure*}[!ht]
    \centering
    \includegraphics[width=\linewidth]{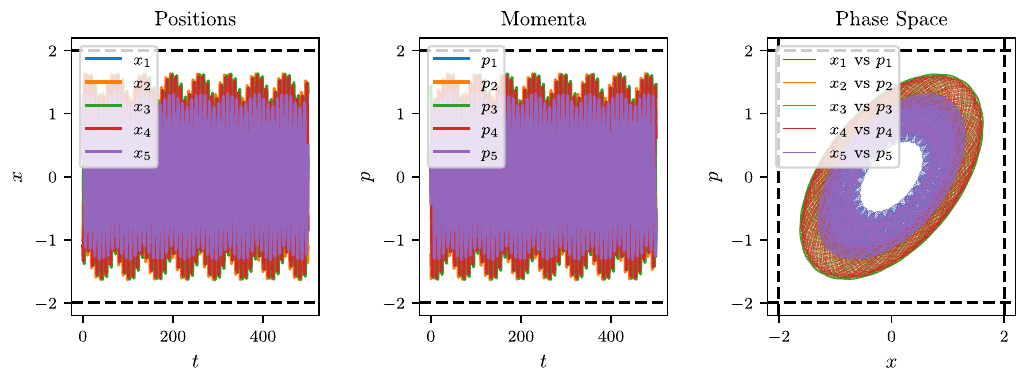}
\caption{Trajectories for $N=5$ with identical global conserved quantities, but otherwise random initial conditions. The bounding ellipse is identical, but trajectories evidently explore different regions of phase space, even at the single particle level. Hence the trajectories are non-ergodic in phase space. The global conserved quantities are taken to be $Q_2=4$, $P_2=4$, $\Pi_{11}=2$, $Q_1=0$, and $P_1=0$.}
    \label{fig:N5}
\end{figure*}

Surprisingly, the ellipse picture generalizes to higher particles.
As can be seen for $N=4$ in \cref{fig:Self_dual}, particles tend to explore phase space, but are constrained within an ellipse.
This ellipse is determined by the global conserved quantities $\mathcal{Q}_2, \mathcal{Q}_1, \mathcal{P}_2, \mathcal{P}_1, \Pi_{11}$, which we have numerically verified for $N > 3$.
The self-dual model is markedly different from the strictly local Machian fractons \cite{AP_NRFractonsPhysRevB.109.054313,AP_manyfractons_PhysRevB.110.024305}: the fractons here do \emph{not} settle into clusters in position space. 
Indeed, Machian clustering in positions is a consequence of the infinite phase space explored.
For this self-dual model, the phase space is bounded by the conservation of $\CQ_2$ and $\CP_2$.
This is intrinsically related to the model not being strictly local: even if $f(R)$ is a decreasing function, particles will be able to influence each other at arbitrary spatial distances.
Nevertheless, true ergodicity is avoided in phase space itself: trajectories maintain quasi-periodic orbits, and refuse to explore the full phase space available within the bounding ellipse. In simulations of $N > 3$ (for example in \cref{fig:N5}) with identical conserved quantities, but otherwise random initial conditions, we always observe exploration of \emph{different} regions of phase space---in other words, ergodicity breaking.

\section{In closing}
In this work, we have classified all general multipole conserving models in classical mechanics, encapsulating and extending previous studies \cite{AP_NRFractonsPhysRevB.109.054313,AP_manyfractons_PhysRevB.110.024305,babbar2025classical}.
On general grounds using the generalized multipole algebra, we have eliminated the possibility of an infinite number of models with multipoles higher than order two in both positions and momenta.
A new self-dual case has emerged, with quasi-periodic orbits and locality features not seen in prior models \cite{AP_manyfractons_PhysRevB.110.024305}.
Extensions of this work would be to develop a full understanding of the self-dual model for more particles and higher dimensions.
We have also presented a construction for Hamiltonians with general momenta multipole conservation laws --- an investigation of their properties is a natural next step.
Quantizing continuum fractons uncovers many unusual behaviors, as we have demonstrated in a recent work~\cite{sadki2025continuumfractonsquantizationbody}; we therefore expect quantizing phase space fractons to also be fruitful.
It would also be interesting to explore lattice models satisfying equivalent conservation laws~\cite{classenhowes2024universalfreezingtransitionsdipoleconserving}.

        \medskip 
    \noindent\emph{Acknowledgments}:  
    AP was supported by the European Research Council under the European Union Horizon 2020 Research and Innovation Programme, Grant Agreement No. 804213-TMCS and the Engineering and Physical Sciences Research Council, Grant number EP/S020527/1. SLS, YS and DPA were supported by a Leverhulme International Professorship, Grant Number LIP-202-014. SLS would also like to acknowledge support by EPSRC via grant EP/X030881/1.
    DPA is grateful for the hospitality of the Physics Department at the University of Oxford, where this work was initiated.  For the purpose of Open Access, the authors have applied a CC BY public copyright license to any Author Accepted Manuscript version arising from this submission. 


\section{End matter}

\subsection{$l = 1, m \geq 1$ Hamiltonians} \label{appendix:non local momenta hamiltonians}
Consider a variation of \cref{eq:old_Ham}.
\begin{equation}
    H = \sum_{\Gamma_{k + 1}} f(R(\{\Bx\nd_\Gamma , \Bp\nd_\Gamma\}))\, g(p_1 - p_2, p_1 - p_3, \dots)\quad,
\end{equation}
where $f$ is a decaying function.
After exchanging $x$ and $p$, $R$ is now:
    \begin{equation}
        R(\Bx\nd_\Gamma , \Bp\nd_\Gamma \}) = \det \begin{pmatrix}
            1 & 1 & \cdots & 1\\
            x_1 &  x_2 & \cdots & x_{k + 1}\\
            p_1 &  p_2 & \cdots & p_{k + 1} \\
            p_1^2 &  p_2^2 & \cdots & p_{k + 1}^2 \\
             & & \vdots & \\
             p_1^{k-1} &  p_2^{k-1} & \cdots & p_{k + 1}^{k-1} 
        \end{pmatrix}\quad. \label{eq:R_multipole_dual}
    \end{equation}
Note that $g$ can be any function, and does not need to be local.
If $g$ is not local in momentum differences, there is no reason to expect this model will cluster in momentum space. 
This is in contrast to the dual $\textbf{m} \geq 1, \textbf{n} = 1$ models studied in \cite{AP_manyfractons_PhysRevB.110.024305}, as spatial locality is imposed as a physical requirement; hence, clustering is observed in positions.

\subsection{$\textbf{m} = 0, \textbf{n} \geq 1$ Hamiltonians} \label{appendix: l0 m geq 1}
Consider, as an example, a variation of \cref{eq:old_Ham}:
\begin{equation}
    H = \sum_{\Gamma_{k + 1}} f(R(\{\Bx\nd_\Gamma , \Bp\nd_\Gamma\})) \sum_a p_a^2\quad,
\end{equation}
where $f$ is again a decaying function.
This Hamiltonian only conserves $\CP_1$ up to $\CP_k$, and does not conserve $\CQ_1$.
$R$ is given by \cref{eq:R_multipole_dual}.

\smallskip
\bibliography{references}

\end{document}